\pdfoutput=1
\documentclass[%
 reprint,
 amsmath,amssymb,
 aps,
]{revtex4-1}

\usepackage{graphicx}
\usepackage{dcolumn}
\usepackage{bm}

\begin{document}


\title{Generalized Analysis of Elastic Wave Dispersion Asymmetries \\in Moving Periodic Media}

\author{M. A. Attarzadeh}
\author{M. Nouh}%
 \email{Corresponding author: mnouh@buffalo.edu}
\affiliation{%
 Mechanical \& Aerospace Engineering, University at Buffalo (SUNY), NY, USA\\}

\date{\today}

\begin{abstract}
This work presents a generalized physical interpretation of unconventional dispersion asymmetries associated moving elastic solids. By shifting the notion from systems with time-variant material fields to physically traveling materials, the newly adopted paradigm provides an eloquent take on the dispersion problem and, in the process, highlights discrepancies between both schemes. Equations governing the motion of an elastic rod with a prescribed moving velocity observed from a stationary reference frame are used to predict propagation patterns and asymmetries in wave velocities induced as a result of the induced linear momentum bias. Three distinct scenarios corresponding to a moving rod with a constant modulus, a spatially varying one, and one that varies in space and time are presented. These cases are utilized to extract and interpret correlations pertaining to directional velocities, dispersion patterns, as well as nature of band gaps between moving periodic media and their stationary counterparts with time-traveling material properties. A linear vertical shear transformation is derived and utilized to thoroughly neutralize the effect of the moving velocity on the resultant dispersion characteristics. Finally, dispersion contours associated with the transient response of a finite moving medium are used to validate the entirety of the presented framework. 
\end{abstract}


\maketitle

\section{Introduction}
The physics of wave dispersion in elastic media have spurred a large of number of research efforts over the past few decades \cite{doyle1989wave,achenbach2012wave, brillouin2003wave,mead1996wave, Hussein2014}. Motivated by their unique wave manipulation capabilities, periodic structures exhibiting tunable band gaps, directional wave guidance, and negative effective densities have culminated in a spurt of research efforts \cite{Ruzzene_beaming, Gonella_directivity, chen2014piezo, huang2009negative, Nouh2015smart, Baravelli2013}. Most recently, in pursuit of new functionalities, novel configurations have been presented as pathways to break elastic wave reciprocity \cite{li2014wave}, onset topologically protected states and edge modes \cite{mousavi2015topologically, pal2017edge}, and realize logical gates and diode-like characteristics in the mechanical domain \cite{bilal2017bistable, li2011diode}. Solids with material fields that vary simultaneously in space and time have been the focus of a number of efforts as a benchmark problem to study non-reciprocity of wave propagation in solids \cite{cassedy1963dispersion,cassedy1967dispersion}. The problem has been investigated in the context of one-dimensional structures using a plane wave expansion method (PWEM) \cite{trainiti2016non} and independently explained via Willis coupling in
the strictly scale-separated homogenization limit in both sub and supersonic regimes \cite{Nassar201710}. Space and time dependent variations of stiffness properties via magnetoelastic materials have been also shown to reproduce similar non-reciprocal patterns \cite{Ansari17}. Bloch-based procedures have been recently generalized for time-dependent discrete phononic lattices \cite{Vila2017363, swinteck2015bulk} and locally resonant metamaterials \cite{nassar2017_meta, attarzadeh2018wave}.

Interest in traveling waves of material properties to realize non-reciprocity dates back to moving photonic crystals which blue-shift and red-shift counter-propagating lights of the same incident frequency due to the Doppler effect \cite{biancalana2007dynamics, wang2013optical}. Motivated by that, spatiotemporal material variations have been utilized as a mechanism to achieve similar behavior in a stationary setup. A conceptually similar interpretation has been proposed in mechanical structures \cite{marconi2017physical}. Such perspectives, though insightful when it comes to understanding wave propagation asymmetry in time-dependent systems, assume a mathematical equivalence between moving structures and stationary ones with moving material fields, which tends to overlook major differences in the dispersion behavior. The focus of this work is to investigate and comprehend wave dispersion in moving solids, i.e. ones that physically travel in space with an arbitrary moving velocity; irrespective of any additional spatial or temporal variations, or lack thereof, of their elastic properties.

\begin{figure*}
\includegraphics[width=\textwidth]{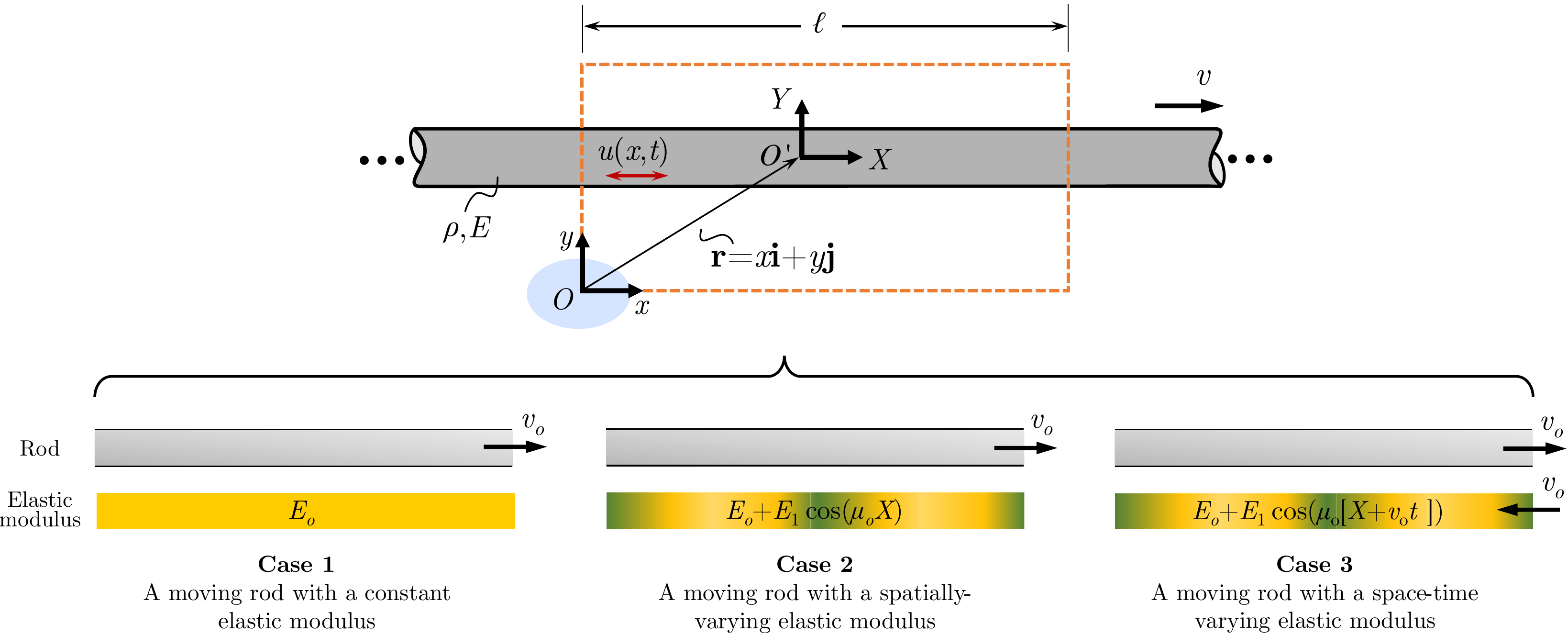}
\centering
\caption{\textit{Top:} Schematic of an elastic rod undergoing axial deformations $u(x,t)$ and moving with a velocity $v$, as observed from a stationary reference ($x$-$y$) and one that is attached to the rod ($X$-$Y$). \textit{Bottom:} Three considered scenarios of a moving rod with a constant velocity $v_o$. Case 1: A constant elastic modulus $E=E_o$. Case 2: A spatially varying elastic modulus $E=E(X)$. Case 3: Elastic modulus $E=E(X,t)$ varies simultaneously in space and time.}
\label{fig:Schematic}
\end{figure*}

The mechanics, stability, and control of moving continua have constituted a classical problem since the 1950s \cite{swope1963vibrations,wickert1990classical,banichuk2010instability,chung1995active}; with applications to subsonic, supersonic, non-accelerating, as well as accelerating velocity regimes \cite{marynowski2014dynamics, ghayesh2012sub, archibald1958vibration, oz1999vibrations}. Most recently, wave propagations in moving elastic media have found new impetus due to the induced linear \cite{ramaccia2017doppler, cummer2014selecting} and angular momentum biases \cite{fleury2014sound} which onset intriguing applications such as magnet-free axisymmetric acoustic circulators \cite{fleury2014sound} and rotating elastic rings \cite{beli2018mechanical}. Despite a few apparent similarities between the wave dispersion characteristics of stationary structures with modulated material properties and those which actually travel, stark differences exist between the dynamics of both categories which have been superficially treated in literature. These include, but are not limited to, differences in group velocities, band gap type, and Brillouin zone shifts, which will be thoroughly highlighted herein. In addition to understanding such differences, a primary goal of this effort is to derive a robust transformation that can neutralize the effect of a non-zero moving velocity on the dispersion diagram of a periodic elastic medium. We start by deriving motion equations pertaining to longitudinal wave propagations in a moving elastic rod as a benchmark example. Following that, dispersion diagrams are extracted for three distinct cases which correspond to constant, spatially-varying, and space-time varying stiffness properties of the moving rod. Finally, the theoretically predicted dispersion characteristics, as well as the asymmetry of group velocities, are validated numerically using a finite realization of the moving rod via a number of finite element simulations.

\section{Physics of a Moving 1D Medium}
We start by deriving the governing equation of motion for a moving elastic rod observed from a stationary reference frame $x$-$y$ (see Fig.~\ref{fig:Schematic}) and undergoing axial vibrations. The rod is assumed to be infinite in length and moving with a prescribed velocity $v$ in the $x$-direction, as depicted in Fig.~\ref{fig:Schematic}. The potential energy $U$ associated with the rod's deformations can be expressed as:

\begin{equation} \label{eq:PE}
U=\frac{1}{2} \int_\ell E \Big( \frac{\partial u}{\partial x} \Big)^2 dx
\end{equation}

\noindent while the non-relativistic kinetic energy $T$ associated with the rod's motion can be written as \cite{shin2004free}:

\begin{equation} \label{eq:KE}
T=\frac{1}{2} \int_\ell \rho \Big( v + \frac{\partial u}{\partial t} + v \frac{\partial u}{\partial x} \Big)^2 dx
\end{equation}

\noindent where $E$ and $\rho$ are the rod's elastic modulus and material density, respectively, and $u(x,t)$ is the rod's axial deformation at any location $x$ and time instant $t$. By writing the system's Lagrangian and employing Hamilton's least action principle, the rod's governing motion equation can be derived as:

\begin{align} \label{eq:EOM}
\nonumber \frac{\partial}{\partial t} \bigg[\rho \Big(v + \frac{\partial u}{\partial t} + v \frac{\partial u}{\partial x} \Big) \bigg] &+
\frac{\partial}{\partial x} \bigg [ \rho v  \Big(v + \frac{\partial u}{\partial t} + v \frac{\partial u}{\partial x} \Big) \bigg]\\
&=\frac{\partial}{\partial x}\Big(E\frac{\partial u}{\partial x}\Big)
\end{align}

In Eq.~(\ref{eq:EOM}), $\rho(x,t)$, $E(x,t)$, and $v=v(x,t)$ (written as $\rho$, $E$, and $v$ for brevity) can generally take any arbitrary functions of space and time. To simplify the analysis, $\rho$ and $v$ will be henceforth assumed constant, i.e. $\rho=\rho_o$ and $v=v_o$. As a result, Eq.~(\ref{eq:EOM}) reduces to \cite{wickert1989energetics}:

\begin{equation} \label{eq:EOM2}
\rho_o \Big (\frac{\partial^2 u}{\partial t^2} + 2v_o \frac{\partial^2 u}{\partial x \partial t}+ v_o^2 \frac{\partial^2 u}{\partial x^2} \Big) = \frac{\partial}{\partial x}\Big(E\frac{\partial u}{\partial x}\Big)
\end{equation}

Eq.~(\ref{eq:EOM2}) is a partial differential equation that describes axial deformations of the elastic rod segment confined within the dashed boundaries in Fig.~\ref{fig:Schematic}. The existence of the second term on the left hand side of the equation indicates the anisotropic (direction-dependent) nature of the problem. Over the next few sections, we will separately investigate three different cases of the moving rod where the elastic modulus $E$ will be assumed constant (Case 1), varying in space similar to a phononic crystal (Case 2), and simultaneously varying in space and time (Case 3). While the rod is continuously moving forward, the focus will be on the segment of the rod falling within the dotted window in Fig.~\ref{fig:Schematic}. For wave propagation purposes, this window is assumed to be sufficiently large such that waves do not reach its boundaries in the process of this analysis. This will enable us to generalize the obtained interpretations and establish some analogies with an equivalent system which comprises a stationary rod with moving material properties, which have been the focus of multiple recent efforts \cite{trainiti2016non, Nassar201710, Ansari17, Vila2017363, swinteck2015bulk, nassar2017_meta, attarzadeh2018wave}.

\section{Wave Dispersion Analysis}

\subsection{Case 1: A moving rod with a constant elastic modulus}

\begin{figure}[h!] 
\includegraphics[width=0.48\textwidth]{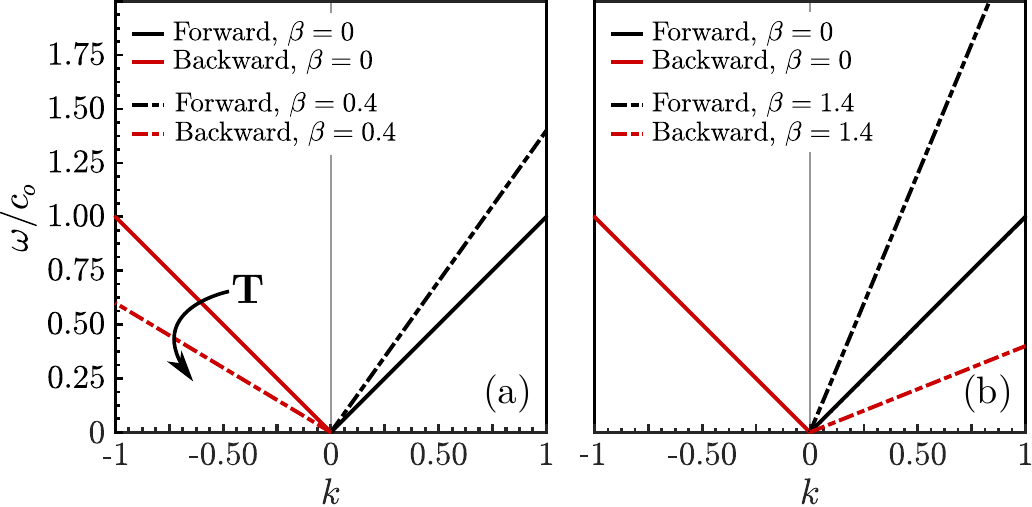}
\centering
\caption{Dispersion diagram of a moving rod with a constant elastic modulus and a relative moving velocity $\beta$: (a) $\beta=0.4$ and (b) $\beta=1.4$. (Dispersion of the stationary rod ($\beta=0$) shown for comparison)}
\label{fig:Unmodulated_Dispersion}
\end{figure}

Studying the moving rod in the absence of any material variations can provide fundamental insights into the effect of the moving velocity on the wave dispersion and propagation characteristics of the elastic medium, which are lacking in the current literature. For non-existent or a slow temporal variation of the elastic modulus, Eq.~(\ref{eq:EOM2}) can be cast in the frequency domain as:

\begin{equation} \label{eq:EOM3}
-\rho_o \omega^2 \hat{u} + 2 i \omega \rho_o v_o \frac{d \hat{u}}{d x}+ \rho_o v_o^2 \frac{d^2 \hat{u}}{d x^2}=\frac{d}{d x}\Big(E\frac{d \hat{u}}{d x}\Big)
\end{equation}

\noindent where $i=\sqrt{-1}$, $\hat{u}$ is the temporal Fourier transformation of $u$ and $\omega$ is the associated frequency. We start by assuming a constant elastic modulus $E=E_o$ and, consequently, a speed of sound $c_o=\sqrt{E_o/\rho_o}$. As a result of the constant modulus, Eq.~(\ref{eq:EOM3}) can be further simplified to:

\begin{equation} \label{eq:EOM4}
(1 - \beta^2)\frac{d^2 \hat{u}}{d x^2} - 2 i \beta\frac{ \omega}{c_o} \frac{d \hat{u}}{d x}+ \frac{\omega^2}{c_o^2} \hat{u}=0
\end{equation}

\noindent where $\beta=v_o/c_o$ is defined as the relative moving velocity. By assuming a harmonic plane wave solution $\hat{u}(x)= Ue^{-ikx}$ with an amplitude $U$ and a wavenumber $k$, the roots of Eq.~(\ref{eq:EOM4})'s characteristic polynomial are shown to be:

\begin{equation} \label{eq:disp_case1}
k(\omega)=\frac{\pm \omega /c_o}{1 \pm \beta}
\end{equation}

\noindent where the wavenumbers corresponding to forward and backward propagations are $k_{F}=\frac{\omega /c_o}{1 + \beta}$ and $k_{B}=\frac{- \omega /c_o}{1 - \beta}$, respectively. Eq.~(\ref{eq:disp_case1}) essentially describes the dispersion relation of a homogeneous moving rod. For a stationary rod (i.e. $\beta=0$), Eq.~(\ref{eq:disp_case1}) yields identical forward and backward dispersion relations which expectedly confirms the elastic reciprocity of the propagating waves. This is, however, no longer the case for $v_o \neq 0$. For a rod moving with a constant velocity the dispersion diagram becomes asymmetric about $k=0$ as a result of the difference in group velocities between the forward and backward traveling waves \cite{chamanara2017optical}, as shown in Fig.~\ref{fig:Unmodulated_Dispersion}. This difference in group velocities is a direct consequence of a \textit{linear momentum bias} caused by the rod's moving velocity $v_o$. Alternatively, the dispersion asymmetry can also be interpreted in light of the Doppler frequency shift effect which is analogous to red-shifting and blue-shifting of light \cite{biancalana2007dynamics}.

Once the rod's velocity exceeds $c_o$ (corresponding to $\beta > 1$), the $k_B$ dispersion branch flips over to the other side indicating that backward traveling waves never reach the observer in such a scenario. The previous phenomenon is displayed in Fig.~\ref{fig:Unmodulated_Dispersion}b. It should be noted here that higher moving velocities risk the possibility of traveling wave discontinuities (reminiscent of sonic booms \cite{Nassar201710}) and time-growing waves that are associated with unstable interactions between the propagating waves and the moving medium \cite{cassedy1963dispersion, cassedy1967dispersion}.

\begin{figure*}
\includegraphics[width=0.95\textwidth]{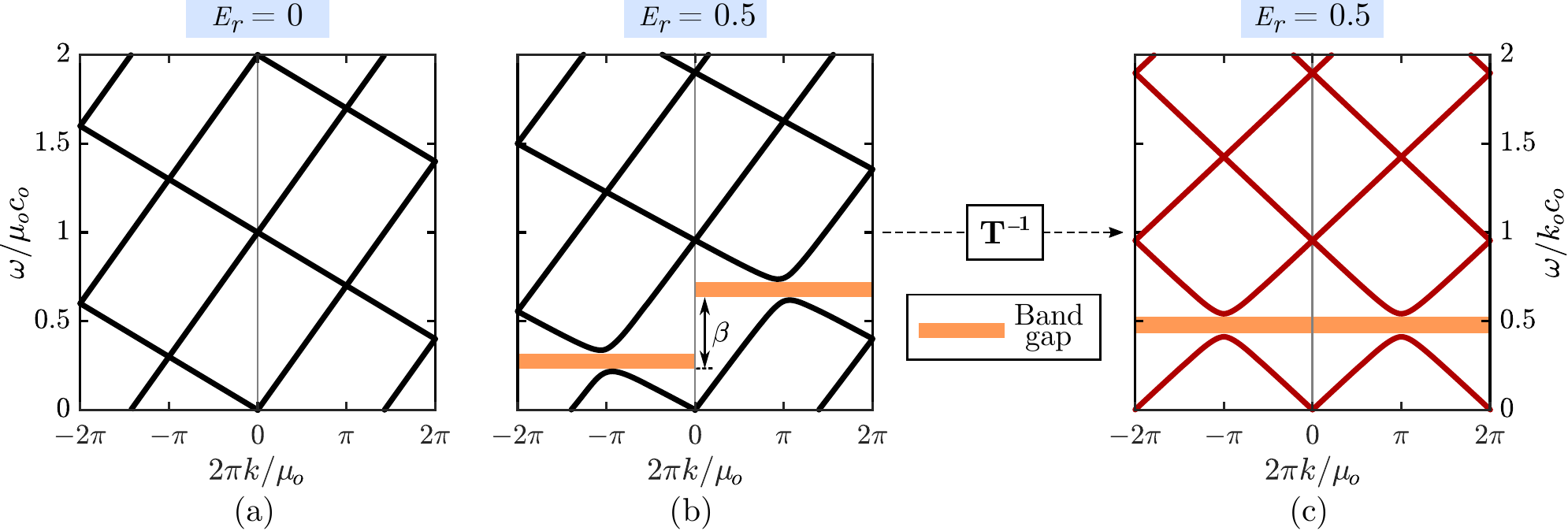}
\centering
\caption{Dispersion diagrams of a moving rod with a relative moving velocity $\beta=0.4$ and (a) $E_r=0$ and (b) $E_r=0.5$. (c) Recovered dispersion diagram of a stationary rod with $E_r=0.5$ upon applying the transformation $\mathbf{T^{-1}}$. (Shaded regions denote frequency band gaps)}
\label{fig:Dispersion_SM_Moving}
\end{figure*}

A closer look at Eq.~(\ref{eq:disp_case1}) and the branches in Fig.~\ref{fig:Unmodulated_Dispersion} reveals that the effect of the non-zero moving velocity on the dispersion behavior can be potentially captured by a linear coordinate transformation. In other words, the dispersion space of the moving medium ($\beta \neq 0$) can be related to that of the stationary one ($\beta=0$) via a transformation matrix $\mathbf{T}$, which can be obtained by rearranging Eq.~(\ref{eq:disp_case1}) in matrix form, as follows:

\begin{equation} \label{eq:in_matrix_form}
\begin{bmatrix}
k\\ \omega/c_o
\end{bmatrix}
_{\mathbf{M}}
= \mathbf{T}
\begin{bmatrix}
k\\ \omega/c_o
\end{bmatrix}
_{\mathbf{S}}
\end{equation}

\noindent where the subscripts $\mathbf{M}$ and $\mathbf{S}$ subscripts denote the moving and stationary rods, respectively. Eq.~(\ref{eq:in_matrix_form}) shows that the intrinsic relation between dispersion spaces in stationary and moving systems is a linear vertical shear transformation with a factor of $\beta$, where $\mathbf{T}$ given by:

\begin{equation} \label{eq:T}
\mathbf{T}=
\begin{bmatrix}
1 & 0 \\
\beta & 1 \\
\end{bmatrix} 
\end{equation}

The inverse of $\mathbf{T}$, i.e. $\mathbf{T^{-1}}$, effectively restores the dispersion diagram of the moving rod to its stationary counterpart. Furthermore, it can be observed that the transformation $\mathbf{T}$ (as well as its inverse) remains well-defined and independent of the frequency $\omega$ for any value of $\beta$. The previous is particularly beneficial for the analysis of unconventional velocity regimes, especially when $\beta \geq 1$. In the following subsections, we will demonstrate how such a geometrical interpretation of the moving velocity effect enables a more intuitive understanding of the interaction between space-time and dispersion domains.

\subsection{Case 2: A moving rod with a spatially-varying elastic modulus}

In this case, we assume an elastic modulus that varies as a function of space such that $E(X)=E_o+E_1 \cos(\mu_o X)$ for any point $X$ in the moving reference frame attached to the rod (see Fig.~\ref{fig:Schematic}). In other words, an observer traveling with the rod would only see a sinusoidal variation of elasticity in space, i.e. a conventional functionally graded phononic crystal (PC) with an elastic modulus that is spatially modulated with a frequency $\mu_o$. In order to utilize the mathematical framework derived in Eq.~(\ref{eq:EOM2}), we need to establish the corresponding elastic modulus variation with respect to the stationary reference frame $x-y$ for the segment of the moving rod within the bounded region in Fig.~\ref{fig:Schematic}. This is given by:

\begin{equation} \label{eq:elastic_variation}
E(x,t)=E_o+E_1 \cos \big(\mu_o [x-v_ot] \big)
\end{equation}

By applying the PWEM, the dispersion characteristics of the moving rod can be obtained. Using the Floquet-Bloch theorem, a harmonic free wave solution with a properly modulated amplitude is given by \cite{cassedy1963dispersion}:

\begin{equation} \label{eq:amp_expansion}
u =e^{i(\omega t-k x)} \sum_{n=-\infty}^{\infty} \tilde{u}_n e^{in\mu_o ( v_o t- x)}
\end{equation}

\noindent where $\tilde{u}_n$ for every $n\in \mathbb{Z}$, is the Fourier coefficient for the amplitude's series expansion. Similarly, $E$ and $\rho_o$ can be expanded as \cite{trainiti2016non}:

\begin{subequations} \label{eq:E_rho_expansion}
\begin{equation}
E=\sum_{m=-\infty}^{\infty} \tilde{E}_m e^{im\mu_o(v_o t-x)}
\end{equation} 
\begin{equation}
\rho_o=\sum_{m=-\infty}^{\infty} \tilde{\rho}_m e^{im\mu_o(v_o t-x)}
\end{equation}
\end{subequations}

\noindent where $\tilde{E}_{m}$ and $\tilde{\rho}_{m}$, with $m \in \mathbb{Z}$, are the $m^{th}$ Fourier coefficients of the expansions, respectively. Upon substituting Eqs.~(\ref{eq:E_rho_expansion}) in (\ref{eq:EOM2}), and exploiting the orthogonality of harmonic functions, the dispersion relation can be expressed as:

\begin{equation} \label{eq:determinant}
\det (\mathbf{\Phi}^{(2)} \omega^2 + \mathbf{\Phi}^{(1)} \omega + \mathbf{\Phi}^{(0)} ) = 0
\end{equation}

\noindent where $\mathbf{\Phi}^{(j)}$, with $j=0,1,2$, are matrices whose $m^{th}$ row and $n^{th}$ column entries are functions of the wavenumber $k$ and are given by:

\begin{equation} \label{eq:phi0}
\phi^{(0)}_{m,n}= v_o^2 k\tilde{\rho}_{m-n} - (k+ m \mu_o)(k+ n \mu_o)\tilde{E}_{m-n}
\end{equation}
\begin{equation} \label{eq:phi1}
\phi^{(1)}_{m,n}=-2v_o k\tilde{\rho}_{m-n}
\end{equation}
and
\begin{equation} \label{eq:phi2}
\phi^{(2)}_{m,n}= \tilde{\rho}_{m-n}
\end{equation}

Taking the assumed variation of properties into account, i.e. a constant density and spatially varying elastic modulus, we can determine the coefficients $\tilde{\rho}_m=\rho_o \delta_{m,o}$ and $\tilde{E}_m=E_o \delta_{m,o} + 0.5E_1 \delta_{m,\pm 1}$, where $\delta_{p,q}$ denotes the Kronecker delta (i.e. $\delta_{p,q}=1$ if $p=q$, and $0$ otherwise). We also define the elastic modulation ratio as $E_r=E_1/E_o$. As $E_r \to 0$, the analysis reduces to the uniform moving rod (Case 1), which is confirmed by Fig.~\ref{fig:Dispersion_SM_Moving}a. For $E_r=0.5$, the dispersion structure opens up phononic (Bragg) band gaps as a result of the periodic variation of elasticity. Furthermore, since the rod is moving with a velocity $v_o=0.4c_o$, the dispersion symmetry is broken as a result of the induced momentum bias causing different forward and backward propagation patterns. The combined effects of these two attributes results in non-reciprocal band gaps, i.e. band gaps that span different frequency ranges depending on the incident wave's direction, as shown in Fig.~\ref{fig:Dispersion_SM_Moving}b.

\begin{figure}[h!] 
\includegraphics[width=0.48\textwidth]{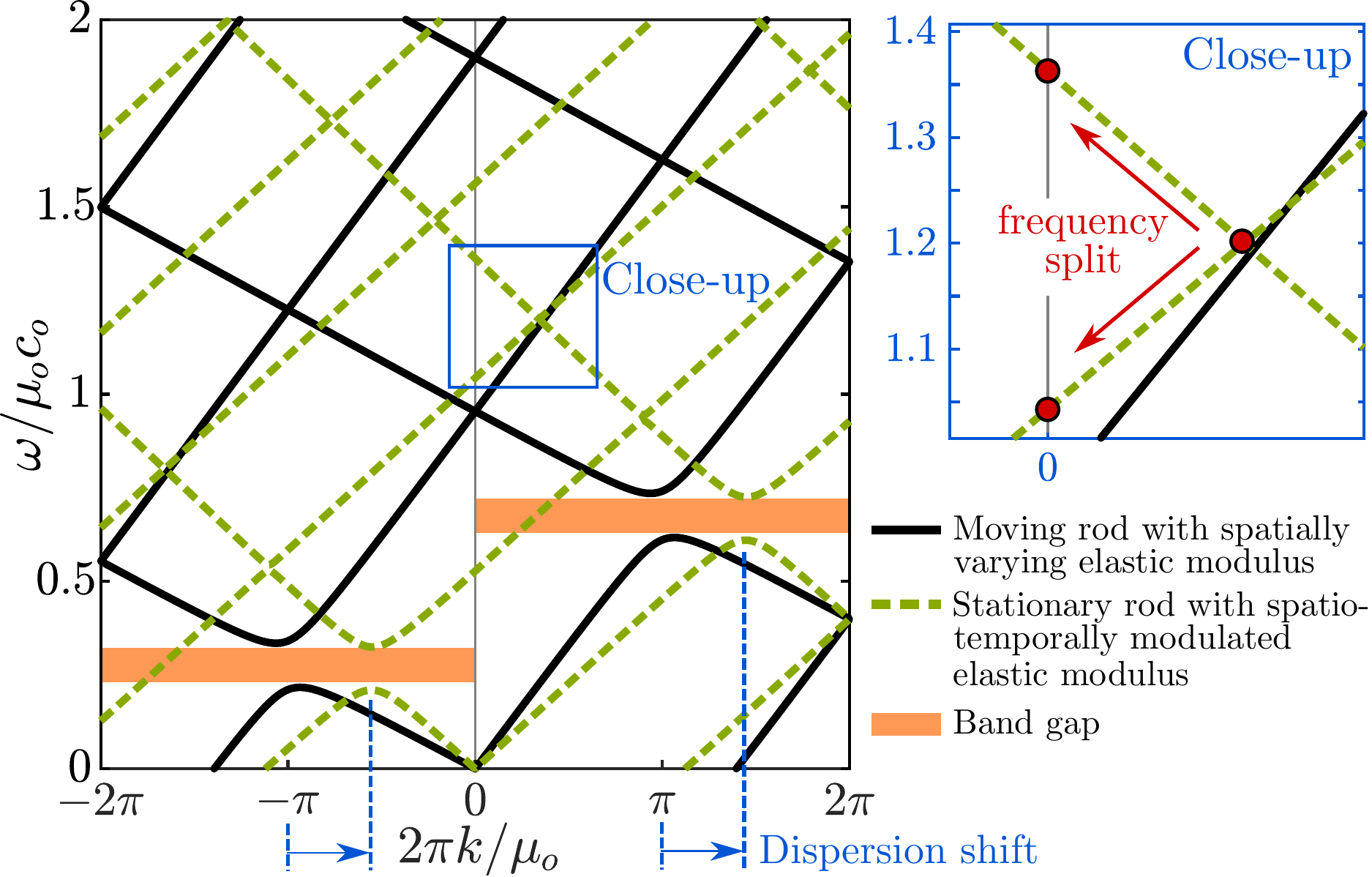}
\centering
\caption{Dispersion differences between a moving rod with a relative moving velocity $\beta=0.4$ and a spatially varying elastic modulus $E$ (\textit{Solid}) and a stationary rod with a spatiotemporally modulated $E$ (\textit{Dashed}). (Shaded regions denote frequency band gaps)}
\label{fig:Moving_Stationary_STM}
\end{figure}

\begin{figure*}
\includegraphics[width=0.95\textwidth]{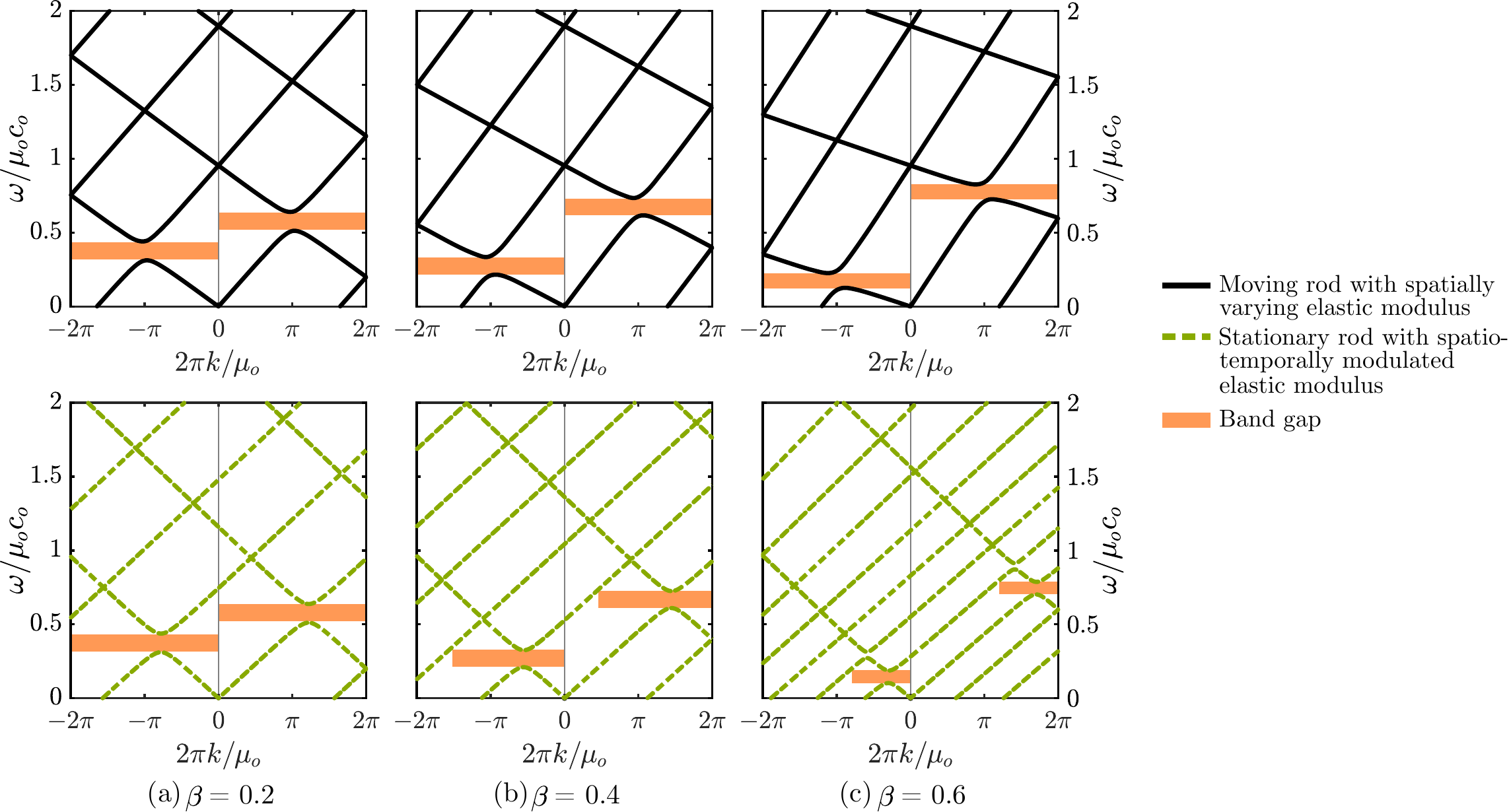}
\centering
\caption{Band gap differences between a moving rod with a relative moving velocity $\beta$ and a spatially varying elastic modulus $E$ (\textit{Top}) and a stationary rod with a spatiotemporally modulated $E$ with a temporal modulation speed $\beta$ (\textit{Bottom}). Comparisons shown for (a) $\beta=0.2$, (b) $\beta=0.4$, and (c) $\beta=0.6$. (Shaded regions denote frequency band gaps)}
\label{fig:Bandgap_Complete}
\end{figure*}

At this stage, a couple of apparent differences can be observed between the non-reciprocal wave dispersion in a moving rod with a spatially varying elastic modulus (i.e. this case) and a stationary rod with a spatiotemporally modulated modulus \cite{trainiti2016non}. In such comparison, which is depicted in Fig.~\ref{fig:Moving_Stationary_STM}, $E_r$ is identical in both cases. Furthermore, the relative velocity of the moving rod $\beta$ is set equal to the modulation speed of the stationary rod's modulus. Although both systems practically share the same non-reciprocal band gap frequency ranges, the shifted dispersion diagram of the stationary rod results in a shift of the Dirac point (where the dispersion bands meet) and, consequently, a frequency split at $k=0$ as can be seen in the close-up shown in Fig.~\ref{fig:Moving_Stationary_STM}. The latter has also been noted in time-dependent periodic lattices \cite{Vila2017363}. On the contrary, despite maintaining the non-reciprocal pattern, such Dirac points in the dispersion diagram of the moving rod consistently take place at integer multiples of $\pm \pi$ of the $x$-axis value. Another stark difference between the two cases is in the nature of the band gap itself. The moving rod exhibits complete (uninterrupted) band gaps which span the entire first Brillouin zone. The complete band gaps always exists irrespective of the value of $\beta$ as can be seen in the upper row of Fig.~\ref{fig:Bandgap_Complete} for $\beta=0.2, 0.4,$ and $0.6$. On the other hand, those taking place in the stationary rod are partial band gaps which diminish in wavenumber coverage as $\beta$ increases, which can be seen in the bottom row of Fig.~\ref{fig:Bandgap_Complete}.

Finally, the effect of the rod's motion on the dispersion behavior can be, once again, nullified by applying the transformation $\mathbf{T^{-1}}$, as shown in Fig.~\ref{fig:Dispersion_SM_Moving}c. As expected, the dispersion diagram shown in Fig.~\ref{fig:Dispersion_SM_Moving}c exhibits perfectly symmetric band gaps of a conventional stationary phononic crystal with $E_r=0.5$.

\subsection{Case 3:  A moving rod with a space-time varying elastic modulus}

\begin{figure}[h!]
\includegraphics[width=0.48\textwidth]{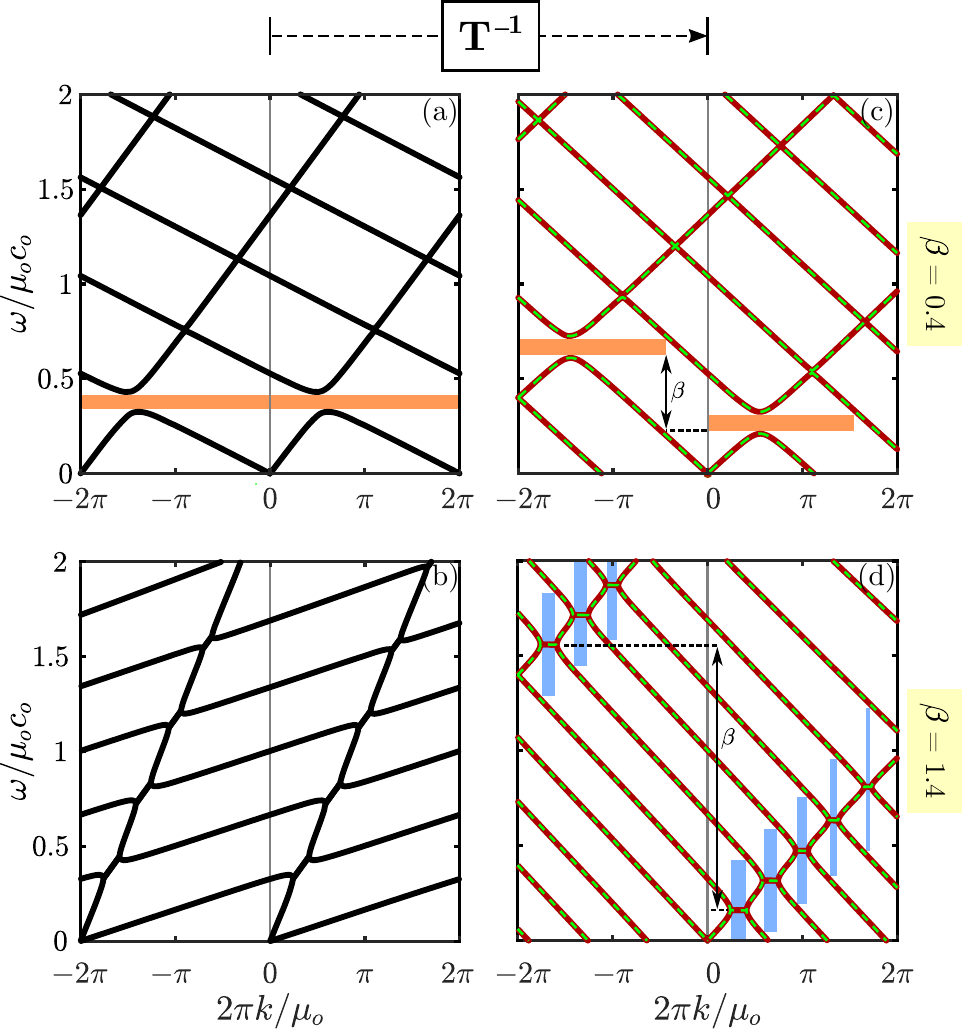}
\centering
\caption {Dispersion diagrams of a moving rod with a space-time varying elastic modulus $E$ with $E_r=0.5$ and a relative moving velocity (a) $\beta=0.4$ and (b) $\beta=1.4$. (c-d) Recovered dispersion diagram of a stationary rod with a space-time varying elastic modulus $E$ upon applying the transformation $\mathbf{T^{-1}}$. Dotted lines are results obtained by using the approach outlined in literature for stationary spatiotemporally modulated systems \cite{trainiti2016non}. (Shaded regions denote band gaps)}
\label{fig:Dispersion_STM_Moving}
\end{figure}

In the last case, we add a final level of complexity to the problem. The rod's elastic modulus is now made to vary simultaneously in space and time with respect to the moving reference frame attached to the rod, and is thus given by $E(X,t)=E_o+E_1 \cos(\mu_o[X+v_ot])$. In addition to that, the rod still travels forward with the moving velocity $v_o$. The expression given for $E$ represents a traveling wave which has a similar speed, but yet an opposite direction to that of the rod's motion. As a result, an observer from the stationary reference frame would only be able to detect the spatial variation of the modulus, which can be alternatively expressed using the time-independent function $E(x)=E_o+E_1\cos(\mu_o x)$. As such, we are able to exploit the frequency domain representation of the governing equation of motion given in Eq.~(\ref{eq:EOM3}). Again, by employing the PWEM, the dispersion relations pertinent to this case can be derived as:

\begin{equation} \label{eq:determinant2}
\det (\mathbf{\Psi}^{(2)} \omega^2 + \mathbf{\Psi}^{(1)} \omega + \mathbf{\Psi}^{(0)} ) = 0
\end{equation}

\noindent where the entries of the matrices $\mathbf{\Psi}^{(j)}$, $j=0,1,2$, are given by:

\begin{equation} \label{eq:phi4}
\psi^{(0)}_{m,n}=(k-n \mu)^2 \tilde{\rho}_{m-n} - (k-m \mu)(k-n \mu)\tilde{E}_{m-n}
\end{equation}

\begin{equation} \label{eq:phi5}
\psi^{(1)}_{m,n}=-2v_o(k-n \mu) \tilde{\rho}_{m-n}
\end{equation}
and
\begin{equation} \label{eq:phi6}
\psi^{(2)}_{m,n}= \tilde{\rho}_{m-n}
\end{equation}

\noindent which comprise the same Fourier coefficients $\tilde{\rho}$ and $\tilde{E}$ used earlier. Dispersion diagrams corresponding to $\beta=0.4$ and $\beta=1.4$ are displayed in Figs.~\ref{fig:Dispersion_STM_Moving}a and b, respectively. These diagrams show two intriguing features: (1) Dispersion diagram asymmetry as a result of the rod's motion can be still observed in both cases, and is evident in the difference in group velocities of the dispersion bands in the long wavelength regime ($k \to 0$). (2) The resultant band gaps are no longer non-reciprocal, i.e. they span the same frequency ranges for forward and backward propagations as depicted by the shaded regions in both figures. The second observation in particular indicates that any band gap non-reciprocity which is created by the \textit{artificial} momentum bias (in this case the imposed spatiotemporal variation of $E$) is effectively canceled by the countering effect of the rod's \textit{actual} momentum (as a result of its non-zero moving velocity $v_o$). As a consequence of these opposite and canceling effects, the resultant dispersion pattern remains asymmetric but no longer exhibits non-reciprocal band gaps.

To further validate the above hypothesis, we again utilize the transformation $\mathbf{T^{-1}}$ to nullify the effect of the rod's motion. The transformed dispersion diagrams (shown in Figs.~\ref{fig:Dispersion_STM_Moving}c and d) are, therefore, representative of a stationary rod with a spatiotemporally modulated elastic modulus; a benchmark case which is established in literature \cite{trainiti2016non, Nassar201710}. The solid lines represent the results obtained by geometrically transforming Figs.~\ref{fig:Dispersion_STM_Moving}a and c using $\mathbf{T^{-1}}$ while the dashed lines represent the results obtained by adopting the mathematical approach used earlier for a stationary rod with a spatiotemporally modulated elastic modulus \cite{trainiti2016non} -- with both results being in perfect agreement. Finally, in the $\beta=1.4$ case (Fig.~\ref{fig:Dispersion_STM_Moving}d), we notice a number of vertical band gaps which correspond to unstable interactions taking place in the super-sonic regime resulting in complex frequencies  akin to a parametric amplifier \cite{cassedy1967dispersion}.

\subsection{Evolution of transfer matrix eigenvalues in moving periodic media}

The previous set of results can also be interpreted in an interesting manner in light of the transfer matrix method (TMM); a commonly used approach in the analysis of periodic systems. In the TMM, a transfer matrix relates the deformation $\hat{u}$ and forcing $\hat{f}$ at one end of a structural segment to its other end. And thus for a periodic system with a number of repetitive partitions subject to an incident excitation, a product of transfer matrices can be used to relay information pertinent to wave propagation in the periodic medium. For a rod with constant elastic modulus $E_o$ which is moving with a constant relative velocity $\beta$, Eq.~(\ref{eq:EOM4}) can be recast into the following state space representation:

\begin{equation} \label{eq:TM0}
\frac{d}{dx}
\begin{bmatrix}
\hat{u}\\ d\hat{u}/dx
\end{bmatrix}
=\begin{bmatrix}
0 & 1 \\ \frac{-\omega^2/c_o^2}{1-\beta^2} & \frac{2i\beta \omega/c_o}{1-\beta^2} 
\end{bmatrix}
\begin{bmatrix}
\hat{u}\\ d\hat{u}/dx
\end{bmatrix}
\end{equation}

\noindent where the strain $d\hat{u}/dx$ can be replaced with $\hat{f}/E_o A$, with $A$ being the cross-sectional area. Eq.~(\ref{eq:TM0}) can be solved for a rod segment of length $l$ which yields: 

\begin{equation} \label{eq:TM0.5}
\begin{bmatrix}
\hat{u}\\ \hat{f}
\end{bmatrix}_{x = l}
=\mathbf{Y}(\beta)
\begin{bmatrix}
\hat{u}\\ \hat{f}
\end{bmatrix}_{x = 0}
\end{equation}

\noindent from which the transfer matrix $\mathbf{Y}(\beta)$ can be derived as:

\begin{align} \label{eq:TM}
\nonumber &\mathbf{Y}(\beta) = \\
&e^{i\beta\Omega l}
\begin{bmatrix}
\cos({\Omega} l)- i\beta \sin({\Omega} l) & \frac{1}{E_o A {\Omega}} \sin({\Omega} l) \\
-(1-\beta^2){E_o A {\Omega}} \sin({\Omega} l) & \cos({\Omega} l)+ i\beta \sin({\Omega} l)
\end{bmatrix} 
\end{align}

\noindent where ${\Omega}=\frac{\omega/c_o}{1-\beta^2}$. For the baseline stationary rod case (where $\beta=0$), $\mathbf{Y}(0)$ in Eq.~(\ref{eq:TM}) yields a \textit{unitary} matrix with a determinant which is equal to one by virtue of the loss-less transmission of information in both forward and backward directions. However, by inspecting $\mathbf{Y}(\beta)$ for a nonzero values of $\beta$, it can be shown that $\det (\mathbf{Y}(\beta)) =e^{2i \beta {\Omega} l}$; an expression which has a magnitude of one and a phase angle of $2 \beta {\Omega} l$. This also indicates that $\mathbf{Y}(\beta)$ of a moving rod with a constant modulus stays unitary irrespective of its speed.

\begin{figure}[h!]
\includegraphics[width=0.48\textwidth]{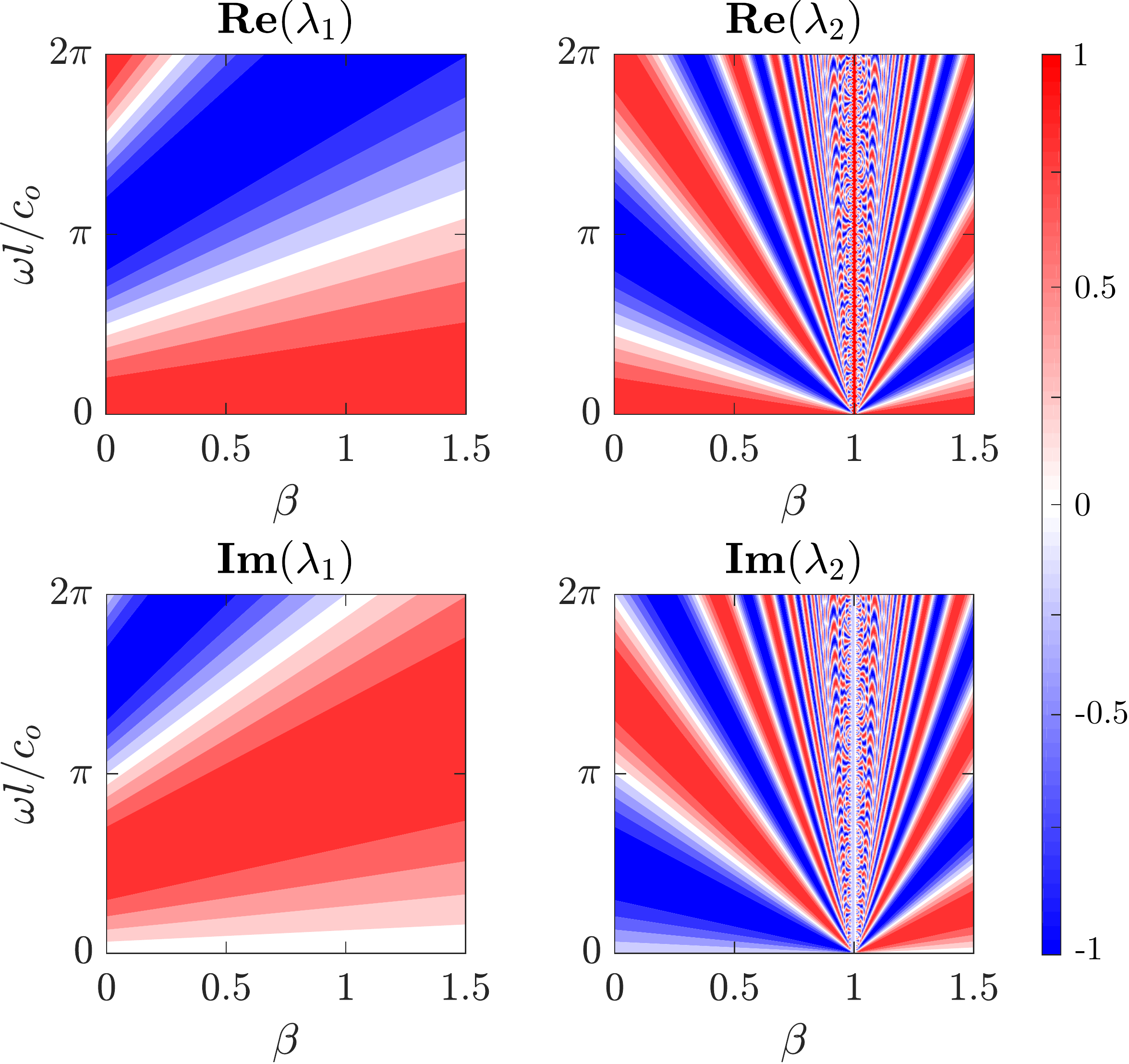}
\centering
\caption {Evolution of the real and imaginary components of the eigenvalues $\lambda_{1,2}$ of the transfer matrix $\mathbf{Y}(\beta)$ for a moving rod with a relative moving velocity $\beta$}
\label{fig:TM_eigenValues}
\end{figure}

Furthermore, it can be shown that the eigenvalues of $\mathbf{Y}(\beta)$ need to satisfy $\lambda_1 \lambda_2 = e^{2i \beta {\Omega} l}$ and $\lambda_1 + \lambda_2 = 2e^{i \beta {\Omega} l} \cos(\Omega l)$, which goes to show that the transfer matrix's eigenvalues evolve as a function of $\beta$. The dependence of both the real and imaginary components of the eigenvalues $\lambda_{1,2}$ on both $\omega$ and $\beta$ is portrayed in Fig.~\ref{fig:TM_eigenValues}. Of interest is the limit of the second eigenvalue $\lambda_2$ ((corresponding to backward traveling waves) which does not exist at the critical moving velocity $\beta=1$. Finally, it is also worth noting that the obtained $\mathbf{Y}(\beta)$ (combined with the transformation $\mathbf{T^{-1}}$) can be further adopted to reproduce the dispersion diagrams of stationary spatiotemporally modulated systems. Unlike the PWEM, the TMM approach renders an exact solution which does not incorporate approximations related to series expansions, the shape of the material modulation, or the number of harmonics included.

\section {Numerical Analysis}

\begin{figure*}
\includegraphics[width=0.95\textwidth]{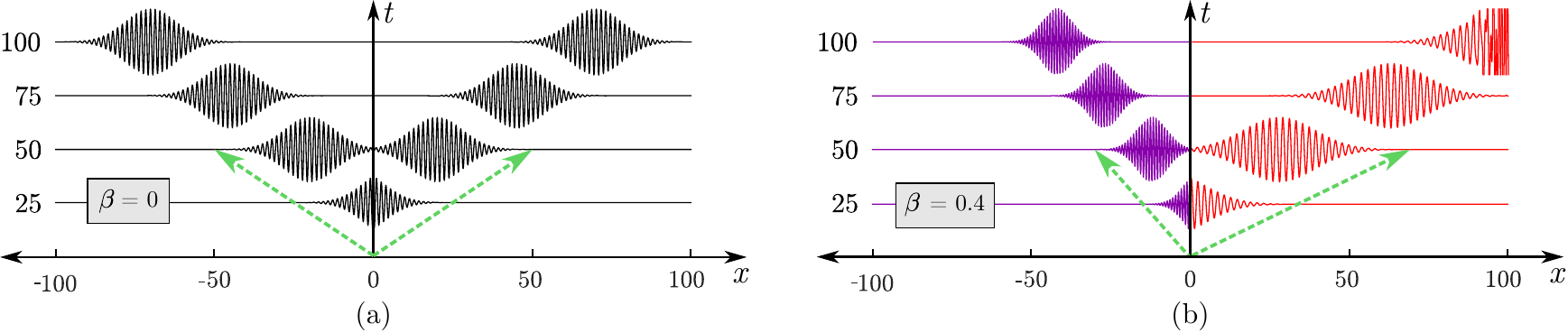}
\centering
\caption{Transient wave propagation in a rod with a constant elastic modulus $E_o$ and a relative moving velocity (a) $\beta=0$ (i.e. stationary) and (b) $\beta=0.4$. (Slope of dotted arrows indicates the group velocity of the chosen wave packet)}
\label{fig:Waterfall}
\end{figure*}

To validate the theoretically obtained dispersion predictions, the transient response of a finite elastic rod of 200 length units divided over 20,000 finite elements is simulated via COMSOL Multiphysics. Starting with a constant elastic modulus $E_o=1$ and an excitation at the rod's mid-span, the transient propagation of a narrow band wave packet along both forward and backward directions of the rod is examined. The wave packet is centered around $\frac{\omega}{c_o \mu_o} = 0.676$ and the simulation spans 100 seconds. As can be seen in Fig.~\ref{fig:Waterfall}a, waves travel in both directions with the same group velocity as predicted earlier in the dispersion diagram of the non-moving rod in Case 1 (Fig.~\ref{fig:Unmodulated_Dispersion}). Once a moving velocity is introduced, the asymmetry of the group velocities in both directions becomes apparent as depicted in Fig.~\ref{fig:Waterfall}b for $\beta=0.4$, and also predicted in Fig.~\ref{fig:Unmodulated_Dispersion}. It should be noted that, in this example, this breakage of wave propagation symmetry takes place in a structure with a time-invariant material field, albeit the structure itself moves with a velocity $v_o$.

\begin{figure}[h!]
\includegraphics[width=0.48\textwidth]{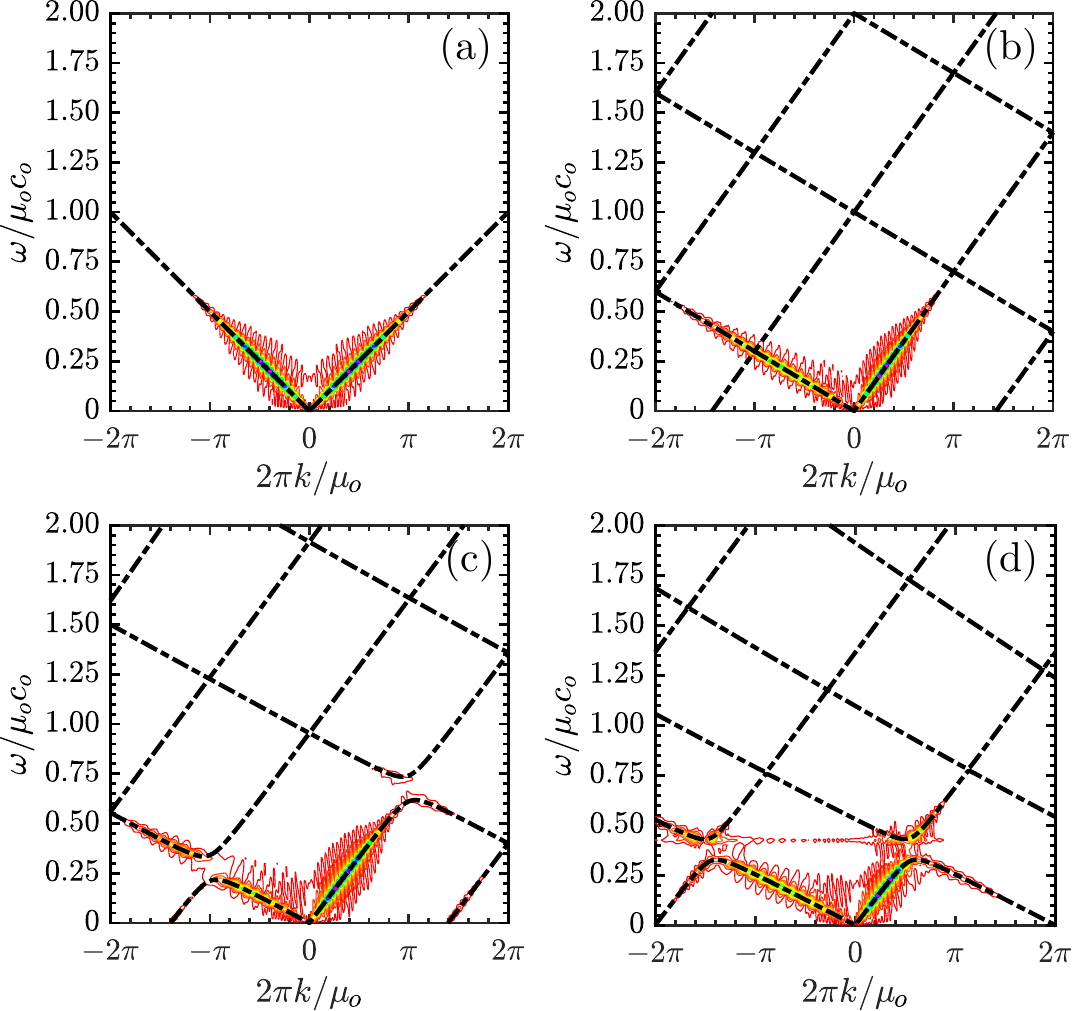}
\centering
\caption{Numerically obtained dispersion contours for (a) a stationary rod with a constant elastic modulus $E_o$, (b) a moving rod with a relative moving velocity $\beta=0.4$ and a constant modulus $E_o$, (c) a moving rod with a relative moving velocity $\beta=0.4$ and a spatially varying modulus ($E_r=0.5$), and (d) a moving rod with a relative moving velocity $\beta=0.4$ and a space-time varying modulus ($E_r=0.5$)}
\label{fig:Numerical}
\end{figure}

The second set of simulations are carried out using a wide band excitation to retrieve the amplitudes of the dispersion contours (versus frequency and wavenumber) of the stationary rod as well as the rods considered in Cases 1, 2, and 3, respectively. The transient longitudinal response of the rod is recorded for the entire length of the structure (comprising 200 unit cells) within a windowed time span which is deemed sufficient for the wave front to reach at least one end of the rod. Dispersion contours are then extracted from the response via a Fourier transform approach \cite{airoldi2011design}. As can be seen in Figs.~\ref{fig:Numerical}a through d, the obtained contours are in strong agreement with the theoretical analysis. thus validating the presented framework for the dispersion analysis of moving elastic periodic media.

\section{Conclusions}

This paper presented a generalized framework and adaptation for the wave dispersion analysis of moving elastic media. By investigating the governing dynamics of an elastic rod with a prescribed moving velocity observed from a stationary frame, a dispersion asymmetry induced as a result of the rod's momentum bias, as well as differences in the associated forward and backward group velocities, can be observed. The analysis studied three distinct cases of the moving rod ranging from a constant elastic modulus, to a spatially varying one, and finally one that varies in space and time independent of the rod's motion. 

A linear vertical shear transformation $\mathbf{T}$ was derived and utilized to add or remove the effect of the moving velocity on the resultant dispersion characteristics. Furthermore, differences as well as correlations -- in dispersion patterns, band gap size, as well as group velocity behavior between moving periodic media and their stationary counterparts with time-traveling material properties were explained. Finally, actual dispersion contours obtained from finite element simulations of the transient response of a moving rod were used to verify and validate the presented mathematical framework. In addition to establishing and understanding these connections between moving and stationary elastic solids, the discussion presented here opens up the possibility of a time invariant analysis of spatiotemporally modulated systems by providing an alternative platform where the calculations and the dispersion predictions can be made and then transformed using the appropriate dispersion transformation. While the design of physically traveling solids is understandably challenging from a practical standpoint, the transformation approach is shown to accurately yield the dispersion behavior of stationary structures with time-varying properties -- albeit without the need to deal with the complexities of time-varying systems. This can be of key significance particularly in the context of synthesis and fabrication of physical realizations of such systems.

%

\end{document}